\documentclass[%
twocolumn,superscriptaddress,
 amsmath,amssymb,
 aps,
prc,
]{revtex4-1}

\usepackage{graphicx}
\usepackage{dcolumn}
\usepackage{bm}
\usepackage{braket}
\usepackage{hyperref}
\usepackage{cancel}
\usepackage{xcolor}
\usepackage{ulem}
\usepackage[T1]{fontenc}

\begin{document}

\title{Non-perturbative calculations of nuclear  matter using in-medium similarity renormalization group}

\author{Xin Zhen}
 \affiliation{
State Key Laboratory of Nuclear Physics and Technology, School of Physics,
Peking University, Beijing 100871, China
}
\author{Rongzhe Hu}
\affiliation{
State Key Laboratory of Nuclear Physics and Technology, School of Physics,
Peking University, Beijing 100871, China
}
\author{Haoyu Shang}
\affiliation{
State Key Laboratory of Nuclear Physics and Technology, School of Physics,
Peking University, Beijing 100871, China
}
\author{Jiawei Chen}
\affiliation{
State Key Laboratory of Nuclear Physics and Technology, School of Physics,
Peking University, Beijing 100871, China
}
\author{J.C. Pei}
\email{peij@pku.edu.cn}
\affiliation{
State Key Laboratory of Nuclear Physics and Technology, School of Physics,
Peking University, Beijing 100871, China
}
\affiliation{
Southern Center for Nuclear-Science Theory (SCNT), Institute of Modern Physics, Chinese Academy of Sciences, Huizhou 516000,  China
}
\author{F.R. Xu}
\affiliation{
State Key Laboratory of Nuclear Physics and Technology, School of Physics,
Peking University, Beijing 100871, China
}
\affiliation{
Southern Center for Nuclear-Science Theory (SCNT), Institute of Modern Physics, Chinese Academy of Sciences, Huizhou 516000,  China
}

\begin{abstract}
The non-perturbative {\it ab initio} calculations of infinite nuclear matter using In-Medium Similarity Renormalization Group (IMSRG) method
is developed in this work, which enables calculations with chiral two and three-nucleon forces at N$^2$LO and N$^3$LO.
Results from the many-body perturbation theory at different orders and coupled-cluster theory are also presented for comparison.
It is shown that different many-body approaches lead to obvious discrepancies with a harder nuclear interaction  for both pure neutron matter
and symmetric nuclear matter.
This work provides a novel alternative infrastructure for future studies of dense nuclear matter and strongly-correlated many-body systems.
\end{abstract}

\maketitle

\section{\label{sec:level1}Introduction}

The homogeneous nuclear matter is a non-trivial strongly-correlated many-body system.
The studies of infinite nuclear matter are especially relevant in modelings of neutron stars in
the era of gravitational-wave astronomy.
The multi-messenger observations of neutron stars
 provide unique
opportunities for constraining the properties of dense nuclear matter \cite{NS_A, NS_B, NS_C, NS_D, NS_E, NS_F, NM14, NM105}.
For example, the recent observation of a massive neutron star of 2.35 solar mass \cite{NS_235}
poses a theoretical challenge, which indicates the appearance of exotic structures in speed of sound
and equation of state (EoS) around two times of nuclear saturation density ($\rho_{\rm sat}$)~\cite{CMJ108}.
At extremely high densities above 40$\rho_{\rm sat}$, the strongly interacting matter can be calculated by the perturbative QCD~\cite{pQCD}.
The key issue is to understand nuclear EoS at intermediate densities around a few times of $\rho_{\rm sat}$,
which is also essential for better
inferences of the first-order phase transition from nuclear matter to quark matter.

The energy of nuclear matter around the saturation density is largely known from finite nuclear properties,
which can be well described by density functional theory based on phenomenological nuclear forces.
However, the slope of symmetry energies has large uncertainties and consequently
 the understanding of neutron-skin thickness is still ambiguous~\cite{nazarewicz}.
 The heavy-ion collision experiments can provide a probe of EoS up to 5 times of $\rho_{\rm sat}$
 but are restricted to nearly symmetric nuclear matter~\cite{heavyion}.
In this context,
it is of fundamental interests to describe nuclear matter beyond the saturation density with
realistic nuclear interactions and {\it ab initio} many-body methods.

The development of modern two-nucleon and three-nucleon interactions from chiral effective field theory (EFT)
provided a consistent theory of nuclear forces rooted in chiral-symmetry breaking of QCD~\cite{CEFTA,CEFTB,CEFTC}.
Advanced  {\it ab initio} calculations based on chiral nuclear forces have
been very successful in descriptions of finite nuclear properties \cite{HH20,NUCLEI_A,NUCLEI_B,KH21},
which can be naturally applied to infinite nuclear matter.
There have been a variety of {\it ab initio} methods for studies of nuclear matter.
For example, the many-body perturbation theory (MBPT) has been extensively applied in a series of works \cite{MBPT1,MBPT2, MBPT3} to calculate EoS of pure neutron matter and symmetric nuclear matter.
The MBPT calculations have been developed up to MBPT4 at 4th order in the momentum space~\cite{MBPT2}.
In addition, the coupled-cluster (CC) theory, which is a non-perturbative method using non-unitary transformations on Hamiltonian has been applied
in studies of nuclear matter \cite{CCladder88, CC89}. The Green function method and Monte Carlo method have also been applied to nuclear matter \cite{AFDMC, CIMC107, SCGFA, SCGFB, Diag110}.
Most of these many-body methods are regarded as post-Hartree-Fock methods to incorporate higher correlations beyond the mean field approximation.

It is expected that non-perturbative calculations are important for strongly-correlated
nuclear matter, particularly, at higher densities.
The purpose of this work is to study nuclear matter from first principles using the
In-Medium Similarity Renormalization Group (IMSRG) method.
IMSRG is a non-perturbative many-body method, which incorporates
many-body correlations efficiently via unitary transformations on the many-body Hamiltonian \cite{IM106,HH16}.
Actually IMSRG provides an alternative non-perturbative method in truncations of many-body operators compared to the coupled-cluster theory.
IMSRG(2) refers to the scheme that all operators are truncated at the normal-ordered two-body level although three-body forces are invoked.
IMSRG(3) is truncated at the normal-ordered three-body level but it is computationally very costly~\cite{IM3A,3F2A,3F2B}.
The coupled-cluster calculations  are usually truncated at the CCSD level that includes single and double particle-hole excitations.
The IMSRG method has been used in ab initio calculations of finite nuclei in recent years~\cite{HH16, NUCLEI_A, NUCLEI_B}, but it
has not been applied in  studies of nuclear matter yet.

\section{Methods}

In this section, we introduce how to implement the IMSRG calculations in momentum spaces based on chiral nuclear forces.

\subsection{Magnus IMSRG(2)}
Firstly, IMSRG is based on a unitary flow transformation~\cite{IM106,HH16} in which the eigenvalue of an operator remains unchanged.
\begin{align}
H(s) = U(s)H(0)U^\dagger(s).
\end{align}
Here the transformation is continuous and $s$ is the flow parameter. One can choose a specific $U$ to obtain a transformed Hamiltonian whose off-diagonal part is as small as possible so that we can solve it in a reduced space. The derivative of $H(s)$ should be
\begin{align}
    \frac{{\rm d}}{{\rm d}s}H(s)=[\eta(s),H(s)],
\end{align}
where $\eta(s)$ is defined as the generator and is related to $U$ as
\begin{align}
    \eta(s)\equiv \frac{{\rm d}U(s)}{{\rm{d}}s}U^\dagger(s)=-\eta^\dagger(s).
\end{align}

The generator is anti-Hermitian because of the Hermicity of $U$. The unitary transformation can be determined by choosing a generator.
The choice of generator $\eta$ depends on which part of Hamiltonian is chosen to be diagonal.
There are different possible types of generators, and  in this work we adopt the White generator~\cite{HH16}, which is simple among different generators and the computing cost is moderate by using the Magnus method~\cite{Magnus92}.
The White generator can be written as
\begin{align}
    \eta^{{\rm White}} = \frac{H^{{\rm od}}}{\Delta}.
\end{align}
Here $H^{{\rm od}}$ is the off-diagonal part we choose, and $\Delta$ is a denominator related to energies. The denominator could be chosen by different energy partitioning such as the M{\o}ller-Plesset scheme,
and the Epstein-Nesbet scheme which we used in this work. Following Ref.~\cite{HH16},  the two-body part of White generator is written as an instance:
\begin{equation}
            \eta_{pp'hh'} \equiv \frac{\Gamma_{pp'hh'}}{f_{p'p'}+f_{pp}-f_{h'h'}-f_{hh}-A_{pp'hh'}},
            \label{eq:eta}
\end{equation}
\begin{eqnarray}
        A_{pp'hh'} &= \Gamma_{pp'pp'}+\Gamma_{hh'hh'}-\Gamma_{phph}\\
        &-\Gamma_{p'h'p'h'}-\Gamma_{ph'ph'}-\Gamma_{p'hp'h}.\nonumber
\end{eqnarray}

Here $p$ and $h$ denote particle states and hole states, which are above or below the Fermi momentum, respectively;  $f$ and $\Gamma$ denote for normal-ordered one-body and two-body Hamiltonian, respectively.
The particle-particle hole-hole configurations are  the off-diagonal part.
The two-body interaction matrix after normal-ordering $\Gamma$ is explicitly suppressed by (\ref{eq:eta}) during the flow evolution.

The IMSRG evolution is a first-order operator differential equation. In order to get precise and stable numerical results, Magnus expansion has been introduced in solving IMSRG equations \cite{Magnus92}. One can rewrite $U$ as
\begin{align}
    U(s) = e^{\Omega(s)}.
\end{align}
By applying the Baker-Campbell-Hausdorff formula, we can get the evolution of $\Omega(s)$ as
\begin{align}
    \frac{{\rm d}}{{\rm d}s}\Omega(s) = \sum_{k=0}^\infty\frac{B_k}{k!}[\Omega(s),\eta(s)]^{(k)}.
\end{align}
Here $B_k$ is the $k$-th Bernoulli number and the $k$-order nested commutator is defined as
\begin{align}
    [\Omega(s),\eta(s)]^{(k)} =
\begin{cases}
\eta(s), & \text{if } k = 0, \\
[\Omega(s),[\Omega(s),\eta(s)]^{(k-1)}], & \text{if } k > 0.
\end{cases}
\end{align}
Therefore, the transformation of Hamiltonian is written as
\begin{align}
    H(s) = e^{\Omega(s)}H(0)e^{-\Omega(s)} = \sum_{k=0}^\infty\frac{1}{k!}[\Omega(s),H(0)]^{(k)}.
\end{align}

Based on this framework, one can perform calculations of the ground state energy of a many-body system non-perturbatively.
It has been pointed out that IMSRG could be regarded as kind of summation of MBPT diagrams to infinite order \cite{HH16}.
This means that IMSRG is a non-perturbative method and could invoke more correlations than MBPT when configuration space is fixed,
which is important as the density increases.

A many-body operator could be given in second-quantized form as
\begin{eqnarray}
        O &= \sum_{pq}T_{pq}a_p^\dagger a_q + \frac{1}{2!}\sum_{pqrs}V_{pqrs}a_p^\dagger a_q^\dagger a_s a_r \nonumber\\&+ \frac{1}{3!}\sum_{pqrstu}W_{pqrstu} a_p^\dagger a_q^\dagger a_r^\dagger a_u a_t a_s+...
        \label{eq:oper}
\end{eqnarray}
In the flow equation, there are commutators of operators which may induce higher-order many-body terms beyond the original Hamiltonian. For instance, the commutation between two-body operators can
induce three-body operators $C^{(3)}$. The commutator in terms of creation and annihilation operators can be written as
\begin{align}
\begin{array}{cll}
    [A^{(2)}, B^{(2)}] & = &C^{(0)} + C^{(1)} +C^{(2)} + \bm{C^{(3)}} \vspace{5pt} \\
    \bm{C^{(3)}_{123456}} &=&\sum_a :a_1^\dagger a_2^\dagger a_4^\dagger a_6 a_5 a_3:  \vspace{5pt} \\
    && \times  (A_{123a}B_{a456}-B_{123a}A_{a456}).
\end{array}
\end{align}

The non-contracted four-body terms vanish after the commutation.
The induced three-body terms can subsequently induce growing many-body terms beyond three-body terms, which means the unitary flow evolution would ultimately contain A-body operators.
Usually, we make a truncation on the flow equation, by dropping all terms higher than a specific order. If we drop $C^{(3)}$ and all induced terms higher than two-body operators, the scheme is called IMSRG(2).
The terms with three-body forces are truncated at the normal-ordered two-body level.
Fortunately the unitary transformations is almost kept with the truncation.
There have been efforts to take back some induced terms that are dropped to improve IMSRG results \cite{3F2A,3F2B}.

\subsection{Interactions in Momentum Space}

For calculations of nuclear matter, it is a natural choice to use plane wave basis with the periodic boundary condition.
One can adopt discretized momentum within a cubic box \cite{CC89}.
Instead, we adopt a sphere in the momentum space to calculate the infinite matter.
The number of nucleons is determined by the lattice points within a definite Fermi sphere.
The number of nucleons is given by:
\begin{align}
    A = g_sg_tN.
\end{align}
Here $g_s, g_t$ reflects the spin-isospin degeneracy and $g_sg_t$ is 4 for SNM and 2 for PNM.
$N=1, 7, 19, 27, 33...$ denotes lattice points within the sphere.
The lattice spacing is determined by the density in calculations, which is similar to Ref.\cite{CC89}.
There could be finite-size effect due to a finite box, but the estimation of finite-size corrections
is very costly.
The finite-size corrections have been verified in terms of kinetic energies and Hartree-Fock energies,
which decrease rapidly with increasing the number of nucleons \cite{CC89}.

The hole-particle definition and normal-order procedure are both related to the Fermi sphere.
As the first-order calculation of IMSRG, the Hartree-Fock (HF) energy in plane wave basis is given by summation as
\begin{align}
    E^{{\rm HF}} = \frac{1}{2}\sum_{i,j}V_{ijij}n_in_j,
\end{align}
where $n_j$ is the occupation number that equals 1 for holes and 0 for particles.
In this work, we mainly adopt $N=33$, corresponding to $4$ shells in momentum spaces for neutron matter.
Such a choice of $N$ parameter has been used in earlier works \cite{CC89, Diag110} and it is relevant to the finite-size correlations.
Afterwards, higher-order correlation energies can be calculated by invoking sufficient particle states beyond Fermi sphere until the convergence is reached.
In this work, the convergence of the correlation energy is generally better than 0.1\%.

Once we built the basis wave functions, we can write down the interaction matrix elements in momentum space.
In most studies, chiral forces are fitted to nucleon-nucleon scattering experiments,
and it is convenient to write the operators in $LSJ$ partial-wave representation.
In some versions of chiral forces, regulators could be different for each partial wave \cite{OPT110,EMN91,EMN96}.
The $LSJ$ representation has been successfully adopted for nuclear matter calculations in the literature~\cite{MBPT2,MBPT3,CDThesis}.
It is complicated to perform transformation from the $LSJ$ representation into the momentum representation.
In this work, the interaction matrix is written in the plane wave basis directly with momentum-dependent regulators.
For tests, the Hartree-Fock energies obtained in the momentum representation have been benchmarked with calculations in the $LSJ$ representation.

Generally, the chiral $NN$ forces could be written as~\cite{CEFTC}:
\begin{eqnarray}
    V_{NN}&&= \mathcal{V}_C + \mathcal{V}_S\bm{\sigma}_1\cdot\bm{\sigma}_2+ \mathcal{V}_T(\bm{\sigma}_1\cdot\bm{q})(\bm{\sigma}_2\cdot\bm{q})\nonumber\\
    &&+\mathcal{V}_{T,k}(\bm{\sigma}_1\cdot\bm{k})(\bm{\sigma}_2\cdot\bm{k})+\mathcal{V}_{LS}(-{\rm i})\bm{S}\cdot(\bm{q}\times\bm{k})\nonumber\\
    &&+\mathcal{V}_{\sigma L}[\bm{\sigma}_1\cdot(\bm{q}\times\bm{k})][\bm{\sigma}_2\cdot(\bm{q}\times\bm{k})].
\end{eqnarray}
Here $\bm{q}$ and $\bm{k}$ are related to momentum of in-state and out-state. The coefficients are defined by:
\begin{align}
    \mathcal{V}_i=V_i^{{\rm cont}} + (V_i+\bm{\tau}_1\cdot\bm{\tau}_2W_i).
\end{align}

The emergence of three-body chiral interactions begins at N$^2$LO in chiral EFT and it is vital for nuclear matter calculation.
It is known that the inclusion of 3$N$ force is essential for reasonable descriptions of saturation properties of symmetric nuclear matter~\cite{NM3N,CDThesis}.
 The 3$N$ force at N$^2$LO  includes three terms:
\begin{align}
V_E = \frac{c_E}{f_\pi^4\Lambda_\chi}\frac{1}{2}\sum_{i\neq j}\bm{\tau}_i\cdot\bm{\tau}_j,
\end{align}
\begin{align}
V_D = -\frac{c_D}{f_\pi^2\Lambda_\chi}\frac{g_A^{}}{8f_\pi^2}\sum_{i\neq j \neq k}\frac{(\bm{\sigma}_i\cdot\bm{q}_j)(\bm{\sigma}_j\cdot\bm{q}_j)(\bm{\tau}_i\cdot\bm{\tau}_j)}{q_j^2+m_\pi^2},
\end{align}
\begin{align}
V_C = \frac{g_A^2}{8f_\pi^2}\sum_{i\neq j \neq k}\frac{(\bm{\sigma}_i\cdot\bm{q}_i)(\bm{\sigma}_k\cdot\bm{q}_k)}{(q_i^2+m_\pi^2)(q_k^2+m_\pi^2)}F_{ijk}.
\end{align}
\begin{eqnarray*}
    F_{ijk} &&= \left(-\frac{4c_1m_\pi^2}{f_\pi^2}+\frac{2c_3}{f_\pi^2}\bm{q}_i\cdot\bm{q}_k\right)\bm{\tau}_i\cdot\bm{\tau}_k
    \\
    &&+\frac{c_4}{f_\pi^2}(\bm{\tau}_i\times\bm{\tau}_k)\cdot\bm{\tau}_j(\bm{q}_i\times\bm{q}_k)\cdot\bm{\sigma}_j.
\end{eqnarray*}

In 3$N$ chiral forces, $f_\pi$, $m_\pi$, $g_A$ and the breakdown scale of chiral EFT $\Lambda_\chi=700\sim 800$ MeV are fixed physical constants.
 The parameters $c_D$, $c_E$ have to be determined by calculations of finite nuclei.
 The parameters $c_i$ are constructed consistently with N$^2$LO two-body forces.
The $c_D$, $c_E$ parameters  are taken from earlier works \cite{CC89,OPT110,NNsat}.

In practical calculations, one should apply regulators on the bare nuclear interactions.
In addition, $NN$ interactions should multiply a relativity factor.
 The detailed expressions of chiral interactions can be found in \cite{EMN91,EMN96,KH21}.
 There are different types of regulators and in this work, the conventional local and nonlocal regulators are employed.
 The expressions of regulators are given in the later part of this work.
 Note that the choice of regulators can impact  calculated EoS of nuclear matter at high densities~\cite{CC89}.

\subsection{Decomposition and Normal-Order}

The interaction matrix elements can be calculated by performing complex matrix products.
 To improve both the speed and accuracy of our calculations, we use automatic plane-wave projection (aPWP) in symbolic Mathematica notebooks
 to deal with  chiral nuclear forces, which is similar to automatic partial-wave decomposition (aPWD) technique \cite{PWD1,PWD2}.
 In aPWP, all potential terms are separated into coefficients and operators.
The expectation values of operators can be calculated in advance.
As an example, for the spin-spin operator $\bm{\sigma}_1\cdot\bm{\sigma}_2$,
the whole spin space is the direct product of the spin spaces of two particles.
Therefore, for two spin-up particles, the $\sigma_{1x}\sigma_{2x}$ component can be calculated as
\begin{align}
    (\bra{\uparrow}\otimes\bra{\uparrow})(\sigma_x\otimes\sigma_x)(\ket{\uparrow}\otimes\ket{\uparrow}) = 0.
\end{align}
For combinations of quantum numbers, all possible values of an operator can be estimated efficiently by the decomposition.
Using this technique, the direct matrix calculations are transformed into case selections in our code, so that calculations become much faster.
 We have separated spin and isospin spaces and finally we obtain four types of decomposition in advance:
\begin{equation}
        \bm{\sigma}_1\cdot\bm{\sigma}_2, \bm{\sigma}\cdot\bm{v}, (\bm{\sigma}_1+\bm{\sigma}_2)\cdot\bm{v}, (\bm{\tau}_1\times\bm{\tau}_2)\cdot\bm{\tau}_3.
\end{equation}
Here $\bm{\sigma}$ and $\bm{\tau}$ are Pauli matrices and $\bm{v}$ is a casual vector. These four kinds of terms cover all operator structures associated with $NN$ and 3$N$ interactions.

Presently, the operators related to three-body forces are truncated at the normal-ordered two-body level.
The full three-body matrix elements are at the scale of $N^6$, compared to $N^4$ for two-body forces.
The treatment of residual three-body forces is a demanding task. Recently IMSRG(3) has been realized for calculations of some light nuclei \cite{IM3A}.
 The normal-ordered two-body Hamiltonian is written as
\begin{align}
    \Tilde{V}_{pqrs} = V_{pqrs}+\sum_{i,j}W_{pqirsj}n_in_j\delta_{ij}.
\end{align}
Here $V$ denotes the original two-body interaction element and $W$ denotes the three-body interaction element, as $\delta$ is the Kronecker symbol.
The expression of zero-body and one-body normal-ordered Hamiltonian also have similar expressions, but have different coefficients \cite{CDThesis, MB09}.
Finally all elements are constructed to be anti-symmetrized.

\subsection{Other Many-Body Methods}

In addition to IMSRG(2), we have built other many-body methods to calculate EoS of nuclear matter for benchmark and comparison purposes.
We have calculated MBPT energy to the 4th order, and the coupled-cluster method is truncated at CCD including double excitations.
All these three methods are based on HF energy as a start.

In MBPT2, the correlation energy beyond HF could be expressed as
\begin{align}
    E^{{\rm MBPT2}}=\frac{1}{4}\sum_{i,j,a,b}\frac{V_{ijab}V_{abij}}{\epsilon_{ij}^{ab}}n_in_j(1-n_a)(1-n_b).
\end{align}
The denominator is defined as
\begin{align}
    \epsilon_{ij}^{ab} = \epsilon_i+\epsilon_j-\epsilon_a-\epsilon _b,
\end{align}
where $\epsilon$ stands for the HF single-particle energy.
The MBPT2 correlation energy can be expressed diagrammatically by one Hugenholtz diagram and the MBPT3 energy contains three diagrams.
Actually, the number of diagrams increases rapidly at higher orders and  brings a challenge in MBPT calculations \cite{ADG,CDThesis,MB09}.
For example, there are $39$ two-body Hugenholtz diagrams in MBPT4.
As mentioned before, IMSRG is non-perturbative and has the capability to sum over diagrams in MBPT to infinite orders.
MBPT2 is the summation of particle-particle hole-hole elements.
In IMSRG(2) evolutions, MBPT2 can be seen as the off-diagonal part of IMSRG(2), reflecting how the suppression and evolution have been achieved.

Another method we use, CC, is also non-perturbative. In contrast to the unitary transformation in IMSRG, CC performs a non-unitary transformation and the evolved Hamiltonian is upper triangular, while the IMSRG Hamiltonian is block diagonal. The analytical formalism of double-excited coupled cluster (CCD) equation  can be found in related literature \cite{CC89,CNP17,MB09}. Since IMSRG relies on unitary transformations, it is convenient to obtain many-body wave functions for future calculations.

\section{Results}

We have built up our program with Fortran 90 from scratch and employed OpenMP for parallel computations.
Currently our code enables calculations of nuclear matter using MBPT, CCD and IMSRG(2).
\subsection{Pure Neutron Matter}
\begin{figure}[htbp]
\includegraphics[width=0.45\textwidth]{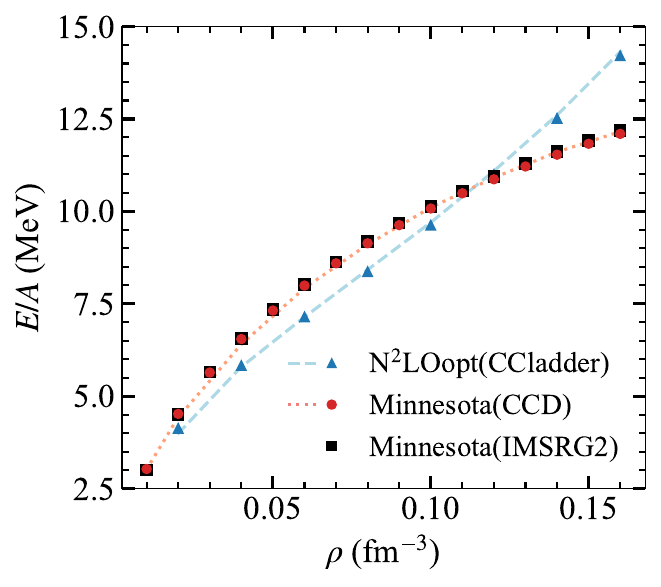}
\caption{\label{fig:Bench} Calculated energy per particle of pure neutron matter using Minnesota and N$^2$LOopt $NN$ potentials for benchmarks.
Dashed lines denote results in \cite{CC89}. }
\end{figure}

Firstly, it is necessary to benchmark our code with existing calculations.
In Fig.~\ref{fig:Bench}, the energy per particle of neutron matter are calculated with Minnesota and chiral potentials.
Minnesota potential \cite{Minne} has a simple form with components like spin-spin and isospin-isospin interactions
and it is convenient for benchmark calculations.
In our calculations, the number of nucleons are taken as $A=66$ for neutron matter and $A=76$ for symmetric matter.
With the Minnesota potential, our IMSRG(2) results agree well with the CCD results in Ref.\cite{CC89}.
The N$^2$LOopt potential~\cite{OPT110} with a cutoff of 500 MeV has also been adopted in our calculations.
With N$^2$LOopt, the CCD calculations with the ladder approximation~\cite{CCladder88} are also benchmarked with Ref.\cite{CC89}, as shown in Fig.~\ref{fig:Bench}.

\begin{figure}[htbp]
\includegraphics[width=0.45\textwidth]{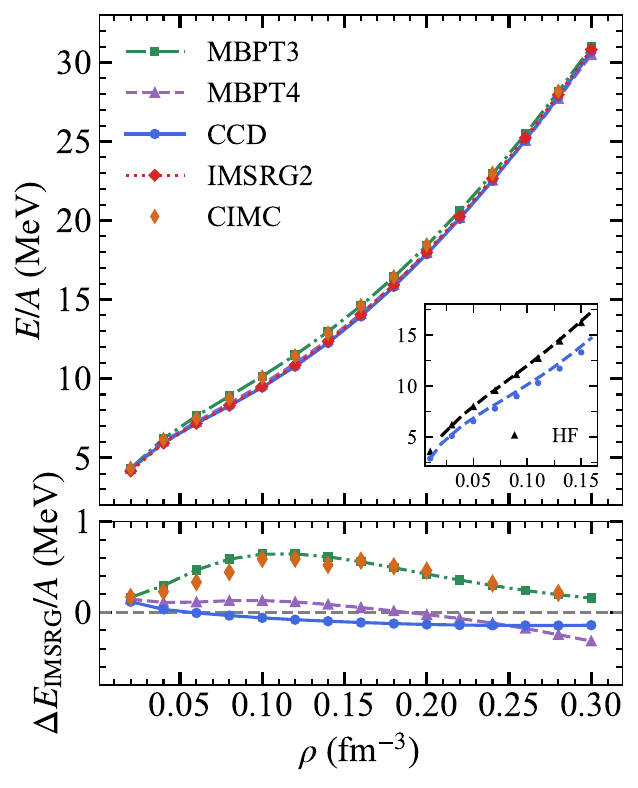}
\caption{\label{fig:NNpnm} Calculated energy per particle of pure neutron matter using N$^2$LOopt $NN$ interactions.
Results from MBPT3, CCD and IMSRG(2), and CIMC from \cite{CIMC107} are compared.
The subfigure shows HF and CCD benchmarks with \cite{CC89}. Deviations of MBPT3, MBPT4, CCD and CIMC to IMSRG2 are shown in the lower figure.}
\end{figure}

In Fig.\ref{fig:NNpnm}, the energy per particle from IMSRG(2) calculations with  N$^2$LOopt are compared with that from different many-body methods.
The results are shown up to 0.3 fm$^{-3}$.
Generally, IMSRG(2) results are very close to CCD results.
The Hartree-Fock and CCD methods have also been benchmarked, as shown in the subfigure.
Our MBPT2 and MBPT3 results agree well with calculations in Ref.\cite{MBPTBench}.
In the intermediate density region below 0.2 fm$^{-3}$, the MBPT3 results are close to that of configuration-interaction Monte Carlo (CIMC)\cite{CIMC107} .
In the low density region below 0.04 fm$^{-3}$, results from different methods are close.
In the high density region above 0.23 fm$^{-3}$, MBPT3 becomes closer to IMSRG(2) and CCD.
MBPT4 results generally are closer to IMSRG(2) compared to MBPT3.
This can be understood that N$^2$LOopt potential is a soft interaction.

\begin{figure}[htbp]
\includegraphics[width=0.45\textwidth]{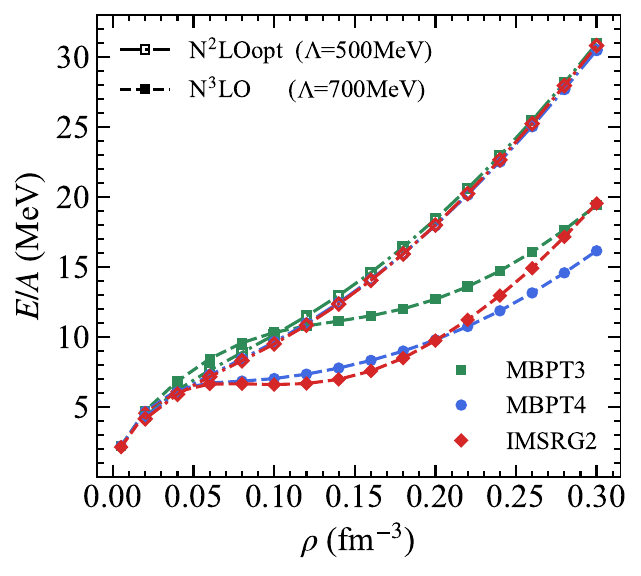}
\caption{\label{fig:NN700} Calculated energy per particle of pure neutron matter using chiral $NN$ forces with different cutoffs.
Results from MBPT3, MBPT4 and IMSRG(2) are compared. }
\end{figure}

It is interesting to compare perturbative and non-perturbative many-body calculations with a harder nuclear interaction.
In chiral nuclear forces,  a higher cutoff usually results in a harder interaction.
In this case, the higher-order many-body correlations are non-negligible and non-perturbative calculations are needed.
In Fig.\ref{fig:NN700}, the energy per particle based on N$^2$LOopt with a cutoff of 500 MeV and  N$^{3}$LO with a cutoff of 700 MeV are shown.
We see that with N$^2$LOopt, MBPT4 results are very close to IMSRG(2), while MBPT3 is slightly different from MBPT4.
For N$^{3}$LO with a cutoff of 700 MeV\cite{SHY110}, the results from different methods have significant discrepancies.
It can be seen that MBPT4 results are relatively close to IMSRG(2) at densities below 0.2 fm$^{-3}$.

\begin{figure}[htbp]
\includegraphics[width=0.45\textwidth]{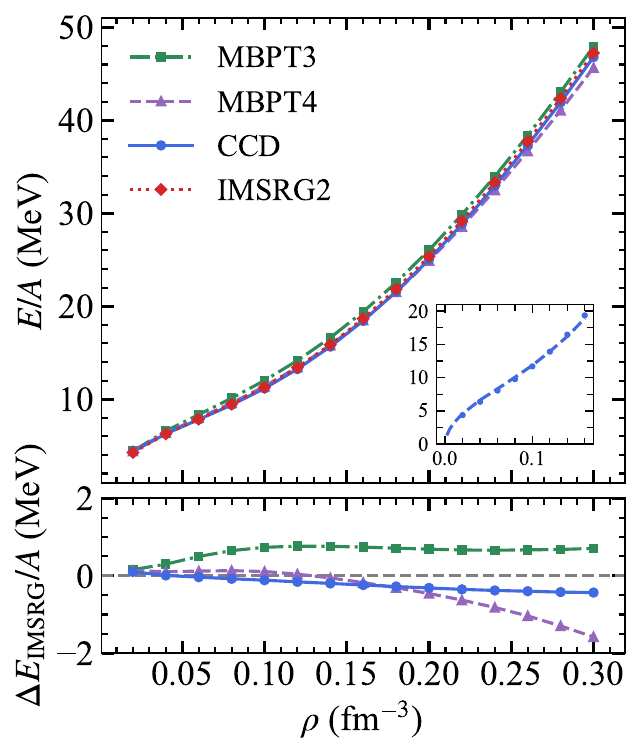}
\caption{\label{fig:3N500} Calculated energy per particle of pure neutron matter using N$^2$LOopt plus nonlocal 3$N$ interactions ($\Lambda$=500MeV).
Results from MBPT3,MBPT4, CCD and IMSRG(2) are compared.
The upper subfigure shows the CCD calculations with three-body forces benchmarked with \cite{CC89}. Deviations of MBPT3, MBPT4 and CCD to IMSRG2 are shown in the lower figure.}

\end{figure}

Three-body nuclear forces are crucial for calculations of nuclear matter.
In this work, we adopted N$^2$LOopt two-nucleon interactions plus three-body interactions that are normal-ordered to two-body level.
For three-body forces, the regulators have both local and nonlocal formalism \cite{OPT110,CC89}.
The nonlocal regulator is written as,
\begin{equation}
    f_{\text{nonlocal}} = \exp\left[-\left(\frac{4p^2+3q^2}{4\Lambda_{3{\rm N}}^2}\right)^n\right].
\end{equation}
Here the variable $p$ and $q$ denote the magnitude of Jacobi momentum of a given three-body state.
For the nonlocal regulator with a cutoff of 500 MeV, we adopt $c_D$=-2 and $c_E$ =-0.791 \cite{CC89}.
The energy per particle of neutron matter are calculated with different methods including three-body forces, as shown in Fig.\ref{fig:3N500}.
With the same nuclear interaction, the results from CCD calculations are benchmarked, as shown in the subfigure.
It can be seen that results from different many-body methods are generally close up to densities at 0.30 fm$^{-3}$.
IMSRG(2) results agree very well with CCD.
The energies from MBPT3 is slightly above that from other three methods.
MBPT4 is close to IMSRG(2) up to 0.2  fm$^{-3}$.
With three-body forces, the EoS of neutron matter becomes harder than that with only two-body forces in Fig.\ref{fig:NNpnm}.

\begin{figure}[htbp]
\includegraphics[width=0.45\textwidth]{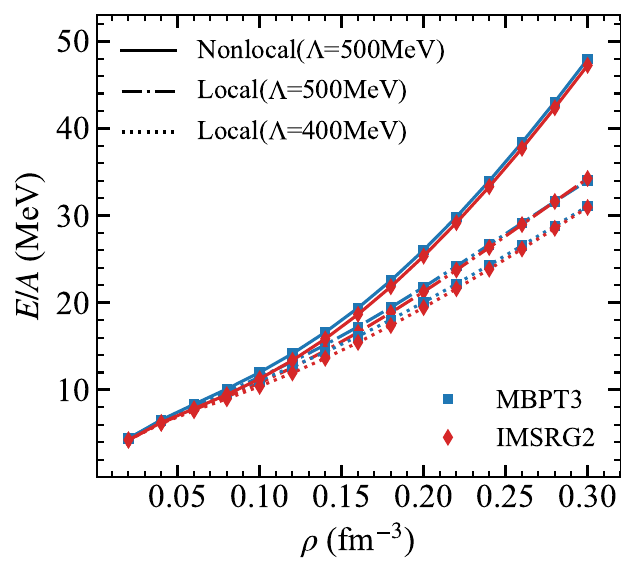}
\caption{\label{fig:3NReg}
Calculated energy per particle of pure neutron matter
 using N$^2$LOopt plus different 3$N$ forces.
 Results from MBPT3 and IMSRG(2) are compared.
Calculations with different 3$N$ nuclear forces with local and nonlocal regulators are shown.
}
\end{figure}

To study the dependence of regulators on the EoS of neutron matter,
the energy per particle are calculated with both local and nonlocal regulators on three-body forces, as shown in Fig.\ref{fig:3NReg}.
The same two-body force  N$^2$LOopt is employed and the three-body forces are dependent on regulators.
The local regulator is written as:
\begin{equation}
    f_{\text{local}} = \prod_{j=1}^{N_j} \exp\left[-\left(\frac{\bm{k}_j'-\bm{k}_j}{\Lambda_{3{\rm N}}}\right)^{2n}\right].
\end{equation}
Here $\bm{k}_j'-\bm{k}_j$ in the exponent factor denote transfer momentum and the index $j$ sums over the transfer momentum  of three nucleons.
For the local regulator with a cutoff of 400 MeV, we adopt $c_D$=-0.39 and $c_E$=-0.27 according to Ref. \cite{CC89}.
For a cutoff of 500 MeV, the parameters are taken as $c_D$ = -0.39 and $c_E$ =-0.389.
It is known that the contact terms have no contributions to neutron matter with a nonlocal regulator.
It can be seen that, with N$^2$LOopt $NN$ plus 3$N$ forces, the EoS with the nonlocal regulator is considerably harder than that with local regulators at higher densities.
We see that MBPT3 results are slightly above IMSRG(2) in three calculations around the saturation density.
However, MBPT3 becomes close to IMSRG(2) around two times of saturation density.
Generally MBPT3 and IMSRG(2) results are close since the cutoff is not high.

\subsection{Symmetric Nuclear Matter}

\begin{figure}[htbp]
\includegraphics[width=0.45\textwidth]{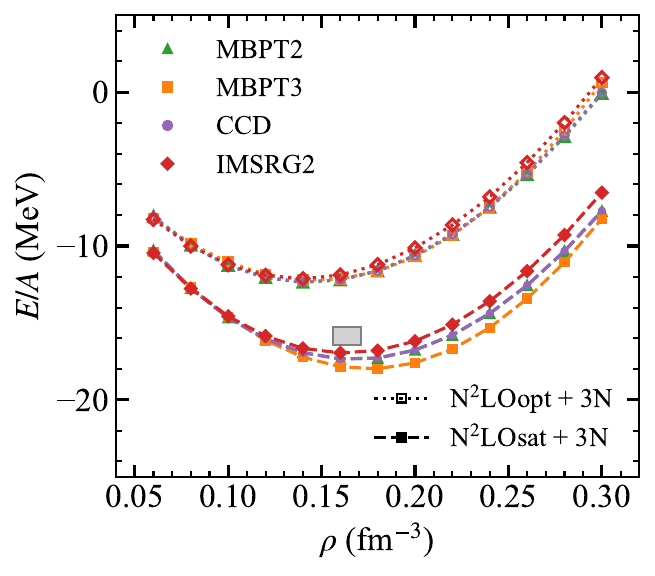}
\caption{\label{fig:3NSNM}
Calculated energy per particle of symmetric nuclear matter using
 N$^2$LOopt and N$^2$LOsat plus nonlocal 3$N$ interactions.
 Results from MBPT2, MBPT3, CCD and IMSRG(2) are compared. }
\end{figure}

For pure neutron matter, we demonstrated that  IMSRG(2) results are close to perturbative calculations, except for
the nuclear interaction is harder with a cutoff of 700 MeV.
It has been pointed out that symmetric nuclear matter is a strongly correlated system, while pure neutron matter is a relatively weakly correlated system~\cite{Diag110}.
It is interesting to study the difference between perturbative and non-perturbative calculations for symmetric nuclear matter.
We adopt both N$^2$LOopt ($\Lambda$=500 MeV) and N$^2$LOsat \cite{NNsat} ($\Lambda$=450 MeV), together with a nonlocal regulator for 3$N$ forces to calculate the EoS of symmetric nuclear matter,  as shown in Fig.\ref{fig:3NSNM}.

 It can be seen that IMSRG(2) using N$^2$LOsat can better describe the empirical saturation point.
Our calculations of symmetric nuclear matter using N$^2$LOsat agree reasonably with recent MBPT3 results \cite{Diag110}.
 We see that different methods have obvious discrepancies for symmetric nuclear matter above $\rho_{\rm sat}$.
 Generally, MBPT2 results are close to CCD results at the normal-ordered two-body level.
  IMSRG(2) is above other methods and the differences become larger at higher densities.
  The discrepancies between different methods are more obvious in calculations with N$^2$LOsat,
 which is a relatively harder potential~\cite{Diag110}.
For N$^2$LOsat,  the growing differences between MBPT2 and MBPT3 indicate that
 MBPT3 is not enough for symmetric nuclear matter at high densities.
 However, MBPT4 calculations of symmetric nuclear matter are extremely costly.
 It has been pointed out that the residual three-body force has non-negligible effects for symmetric matter\cite{CC89},
 which will be our next step.

\section{Discussions}

In this work we have presented the development of the non-perturbative {\it ab initio} calculations of nuclear matter using In-Medium Similarity Renormalization Group method.
IMSRG is an attractive many-body method based on continuous unitary transformations to include infinite summation of MBPT diagrams.
IMSRG method has been applied in studies of finite nuclei in the literature, but has not been used for infinite nuclear matter yet.
Our code is built from scratch and have been benchmarked with coupled-cluster calculations.
The present work IMSRG(2) has been adapted to use chiral two-body forces and three-body forces truncated at the normal-ordered two-body level.

The code enables calculations of IMSRG(2),  MBPT to 4th order and CCD based on the same Hartree-Fock start.
The calculations with different methods have been compared.
The energy per particle from different methods are close for pure neutron matter using two-body forces N$^2$LOopt, which is a soft interaction.
In this case, MBPT4 is closer to IMSRG and CCD results compared to MBPT3.
The situation is the similar for calculations with three-body forces with a cutoff of 500 MeV.
However, with a harder N$^3$LO nuclear force with a cutoff of 700 MeV, there is significant discrepancies between non-perturbative and perturbative calculations, in particular
at higher density region.
Similarly, there is obvious discrepancies between different methods for symmetric nuclear matter with a harder nuclear interaction N$^2$LOsat \cite{NNsat}.
At the moment, our method is truncated at IMSRG(2) in the plane wave basis.
It is costly to perform higher-order IMSRG calculations such as IMSRG(3) and IMSRG(3f2), which have been realized recently for finite nuclei~\cite{3F2A,3F2B}.
It is our next step to extend our calculations towards IMSRG(3) to treat harder interactions and residual three-body interactions.
Such calculations should be insightful for studying EoS of dense nuclear matter.

\acknowledgments
We are grateful for useful discussions with B.S. Hu, W.G. Jiang.
 This work was supported by  the
 National Key R$\&$D Program of China (Grant No.2023YFE0101500, 2023YFA1606403, 2024YFA1610900),
  the National Natural Science Foundation of China under Grants No.12475118, 12335007, 12035001.
We also acknowledge the funding support from the State Key Laboratory of Nuclear Physics and Technology, Peking University (No. NPT2023ZX01).

\end{document}